\documentclass[12pt]{article}

\usepackage{graphicx}
\usepackage{epsfig}

\begin{document}
\begin{center}
\bf PROTON AND $\Lambda$-HYPERON PRODUCTION \\
IN NUCLEUS-NUCLEUS COLLISIONS
\vspace{.2cm}

G.H. Arakelyan$^1$, C. Merino$^2$, and Yu.M. Shabelski$^{3}$ \\

\vspace{.5cm}
$^1$A.Alikhanyan National Scientific Laboratory \\
(Yerevan Physics Institut)\\
Yerevan, 0036, Armenia\\
E-mail: argev@mail.yerphi.am\\
\vspace{0.1cm}

$^2$Departamento de F\'\i sica de Part\'\i culas, Facultade de F\'\i sica\\
and Instituto Galego de F\'\i sica de Altas Enerx\'\i as (IGFAE)\\
Universidade de Santiago de Compostela\\
15782 Santiago de Compostela\\
Galiza-Spain\\
%Santiago de Compostela 15782, \\
%Galiza, Spain \\
E-mail: merino@fpaxp1.usc.es \\
\vspace{0.1cm}

$^{3}$Petersburg Nuclear Physics Institute\\
NCR Kurchatov Institute\\
Gatchina, St.Petersburg 188350, Russia\\
E-mail: shabelsk@thd.pnpi.spb.ru
\vskip 0.9 truecm

%{\it Extende variant of a talk given at HSQCD Conf., St.Petersburg, 5-9
%July 2010}

\vspace{1.2cm}

{\bf Abstract}
\end{center}

The experimental data on net proton and net $\Lambda$-hyperon spectra obtained
by the NA35 Collaboration, as well as the inclusive densities of $\Lambda$
and $\bar{\Lambda}$ obtained by NA49, NA57, and STAR collaborations,
are compared with the predictions of the Quark-Gluon String Model. 
The contributions of String Junction diffusion, interactions with nuclear
clusters, and the inelastic screening corrections are accounted for. The level 
of numerical agreement of the calculations with the experimental data is of about 
20$-$30\%. The predictions for LHC are also presented. 

\vskip 1.5cm


\newpage

%
%{\Large \bf

\section{Introduction}

The Quark-Gluon String Model (QGSM) \cite{KTM} is based on the Dual
Topological Unitarization, Regge phenomenology, and nonperturbative
notions of QCD. This model is successfully used for the description of
multiparticle production processes in hadron-hadron \cite{KaPi,Sh,AMPS,MPS},
hadron-nucleus \cite{KTMS,Sh1}, and nucleus-nucleus \cite{Sha,Shab,JDDS}
collisions. In particular, the rapidity dependence of inclusive densities
of different secondaries ($\pi^{\pm}$, $K^{\pm}$, $p$, and $\bar{p}$)
produced in Pb-Pb collisions at 158 GeV/c per nucleon were reasonably
described in ref. \cite{JDDS}. In the present paper we consider the yields
of $p$ and  $\bar{p}$, as
well as $\Lambda$ and $\bar{\Lambda}$, produced in the collisions of
different nuclei at CERN SpS and RHIC energies.

In the QGSM high energy interactions are considered as proceeding via
the exchange of one or several Pomerons, and all elastic and inelastic
processes result from cutting through or between Pomerons \cite{AGK}.
Inclusive spectra of hadrons are related to the corresponding fragmentation
functions of quarks and diquarks, which are constructed using the Reggeon
counting rules \cite{Kai}.

In the case of interaction with nuclear target, the Multiple Scattering
Theory (Gribov-Glauber Theory) is used. It allows to consider the
interaction with nucleus as the superposition of interactions with
different numbers of target nucleons \cite{Sh3,BT,Weis,Jar}.

Also in the case of nucleus-nucleus collisions the Multiple Scattering Theory
allows to consider the interaction as the superposition of separate
nucleon-nucleon interactions. Though in this case the analytical
summation of all the diagrams is impossible~\cite{BSh}, the signiificant classes
of diagrams can be analytically summed up in the so-called rigid target
approximation \cite{Alk} which is used in the present paper.

The significant differences in the yields of baryons and antibaryons in the central
(midrapidity) region are present even at high energies. This effect can be explained
\cite{AMPS,MPS,ACKS,BS,AMS,Olga,MPS1,MRS} in QGSM by the special structure of
baryons consisting of three valence quarks together with a special
configuration of gluon field, called String Junction \cite{Artru,IOT,RV,Khar}.

One additional contribution comes from the coherent interaction of a projectile with
multiquark clusters inside the nuclei. The existance of these interactions is
confirmed by the presence, with not such a small probability, of a cumulative
effect \cite{SF}.

These contributions were incorporated into the QGSM in
\cite{Efr,Efre}, and they allow to describe a number of experimental facts.

%used in the framework of the QGSM in
%\cite{Efr,Efre} and it is similar to the Monte Carlo String Fusion Model
%\cite{Paj}.

At very high energies the contribution of the enhancement Reggeon diagrams
becomes important, leading to a new phenomenological effect, the supression
of the inclusive density of secondaries \cite{CKTr} in the central (midrapidity) 
region.

In this paper we present the description of $p$, $\bar{p}$,
$\Lambda$, and $\bar{\Lambda}$ production on nuclear targets at CERN SpS and
RHIC energies.

\section{Baryon/antibaryon asymmetry in the QGSM}

\subsection{General approach}

The QGSM \cite{KTM,KaPi,Sh} allows one to make
quantitative predictions for different features of multiparticle production,
in particular, for the inclusive densities of different secondaries, both in
the central and in the beam fragmentation regions.

In QGSM, each exchanged Pomeron corresponds to a cylindrical diagram, and thus, 
when cutting one Pomeron, two showers of secondaries are produced (see Fig.~1 a,b).
\begin{figure}[htb]
\centering
\vskip -1.cm
\includegraphics[width=.4\hsize]{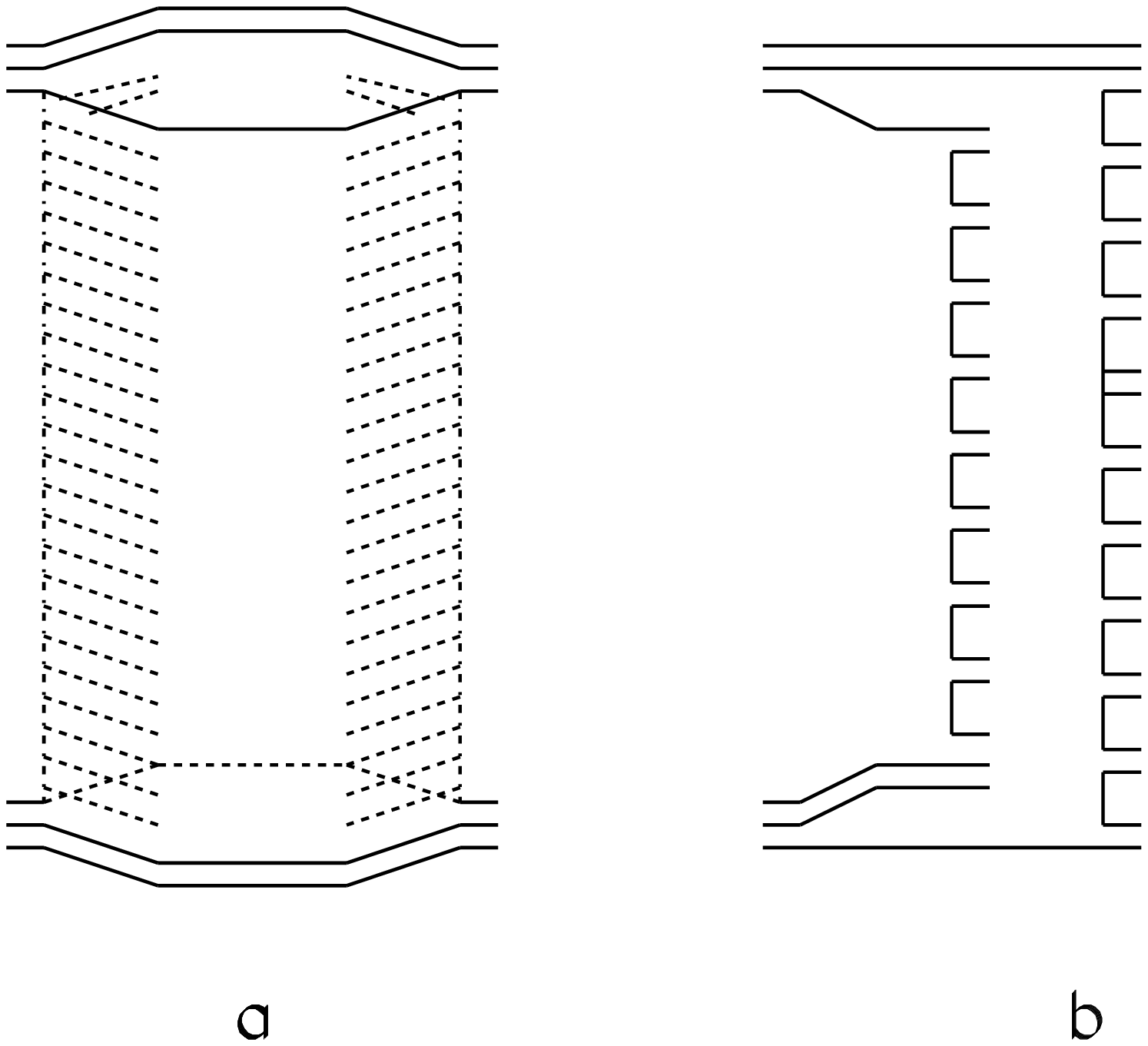}
\vskip -1.7cm
\includegraphics[width=.45\hsize]{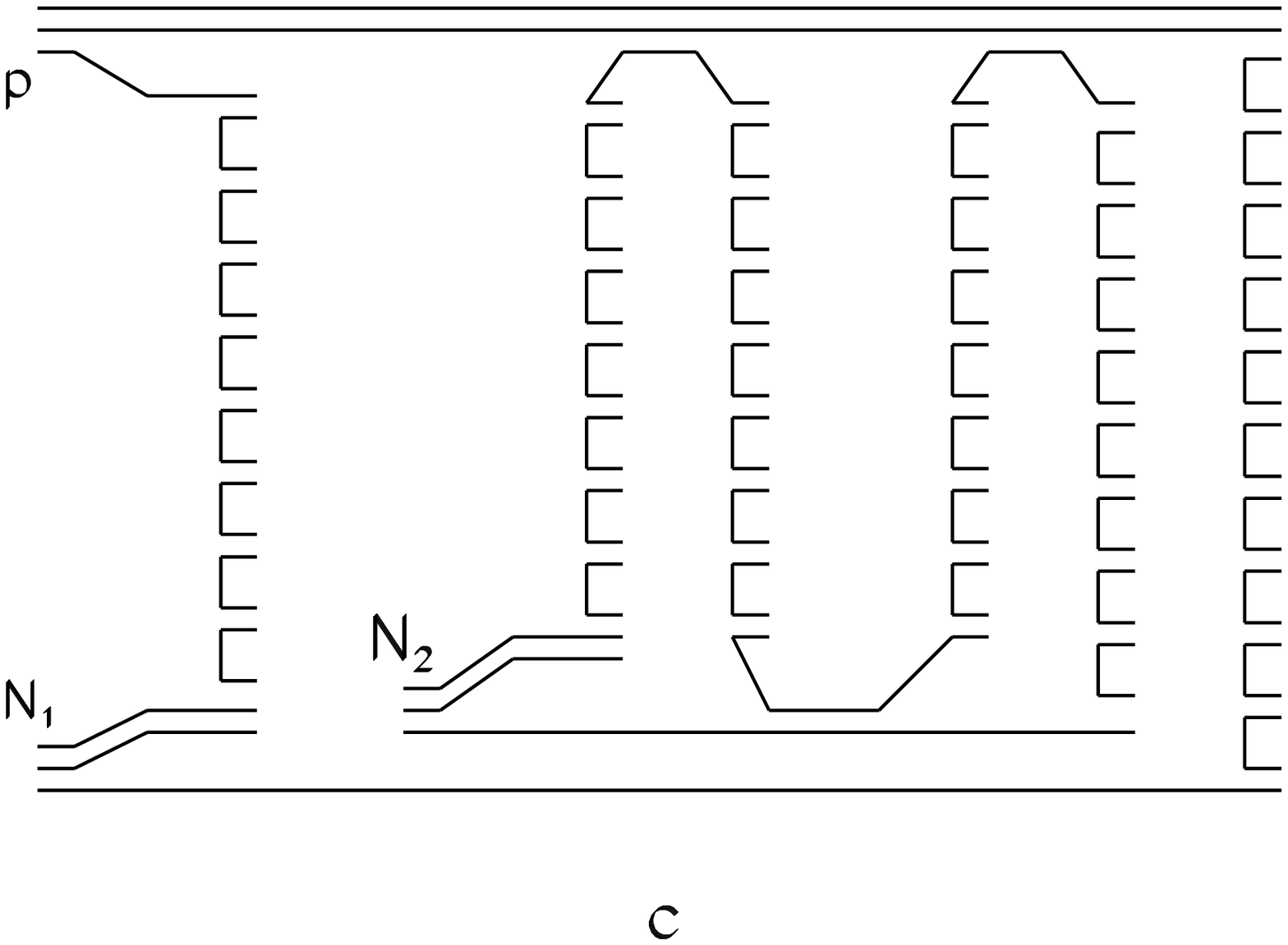}
\vskip -.8cm
\caption{\footnotesize
(a) Cylindrical diagram representing a Pomeron exchange within the DTU
classification (quarks are shown by solid lines); (b) One cut of the cylindrical
diagram corresponding to the single-Pomeron exchange contribution in inelastic
$pp$ scattering; (c) One of the diagrams for the inelastic interaction of one
incident proton with two target nucleons $N_1$ and $N_2$ in a $pA$ collision.}
\end{figure}

The inclusive spectrum of a secondary hadron $h$ is then determined by the
convolution of the diquark, valence quark, and sea quark distributions,
$u(x,n)$, in the incident particles, with the fragmentation functions, $G^h(z)$,
of quarks and diquarks into the secondary hadron $h$. Both the distributions
and the fragmentation functions are constructed using the Reggeon counting rules.

In particular, in the case of $n > 1$, i.e. in the case of multipomeron
exchange, the distributions of valence quarks and diquarks are softened due to
the appearance of a sea quark contribution. There is some freedom~\cite{KTMS}
in how to account for this effect. In principle, the valence and sea quarks can
depend on $n$ in a different manner, for example:
\begin{eqnarray}
u_{uu}(x,n) & = & C_{uu}\cdot x^{\alpha_R-2\alpha_B+1}
\cdot(1-x)^{-\alpha_R+m_1} \nonumber \\ \nonumber
u_{ud}(x,n) & = & C_{ud}\cdot x^{\alpha_R-2\alpha_B}\cdot(1-x)^{-\alpha_R+m_2}
\\ \nonumber
u_u(x,n) & = & C_u\cdot x^{-\alpha_R}\cdot(1-x)^{\alpha_R-2\alpha_B+m_2}
\\ \nonumber
u_d(x,n) & = & C_d\cdot x^{-\alpha_R}\cdot(1-x)^{\alpha_R-2\alpha_B+1+m_1}
 \\
u_s(x,n) & = & C_d\cdot x^{-\alpha_R}\cdot(1-x)^{\alpha_R-2\alpha_B+(n-1)+d} \;,
\end{eqnarray}
where every distribution $u_i(x,n)$ is normalized to unity, $d$ is a
parameter, and the values of  $m_1$ and $m_2$ can be found from the conditions:

\begin{eqnarray}
\langle x_{uu} \rangle + \langle x_d \rangle + 2(n-1)\cdot \langle x_s \rangle
&=& 1 \; \nonumber \\
\langle x_{ud} \rangle + \langle x_u \rangle + 2(n-1)\cdot \langle x_s \rangle
&=& 1 \;.
\end{eqnarray}

The details of the model are presented in~\cite{KTM,KaPi,Sh,ACKS}. The
averaged number of exchanged Pomerons $\langle n \rangle_{pp}$ slowly
increase with the energy. The Pomeron parameters have been taken from \cite{Sh}.

For a nucleon target, the inclusive rapidity, $y$, or Feynman-$x$, $x_F$,
spectrum of a secondary hadron $h$ has the form~\cite{KTM}:
\begin{equation}
\frac{dn}{dy}\ = \
\frac{x_E}{\sigma_{inel}}\cdot \frac{d\sigma}{dx_F}\ = 
\sum_{n=1}^\infty w_n\cdot\phi_n^h (x) + w_D \cdot\phi_D^h (x) \ ,
\end{equation}
where the functions $\phi_{n}^{h}(x)$ determine the contribution of diagrams
with $n$ cut Pomerons, $w_n$ is the relative weight of this diagram, 
and the term $w_D \cdot\phi_D^h (x)$ accounts for the contribution of diffraction
dissociation processes.

In the case of $pp$ collisions:
\begin{equation}
\phi_n^{h}(x) = f_{qq}^{h}(x_{+},n) \cdot f_{q}^{h}(x_{-},n) +
f_{q}^{h}(x_{+},n) \cdot f_{qq}^{h}(x_{-},n) +
2(n-1)\cdot f_{s}^{h}(x_{+},n) \cdot f_{s}^{h}(x_{-},n)\ \  ,
\end{equation}

\begin{equation}
x_{\pm} = \frac{1}{2}[\sqrt{4m_{T}^{2}/s+x^{2}}\pm{x}]\ \ ,
\end{equation}
where $f_{qq}$, $f_{q}$, and $f_{s}$ correspond to the contributions
of diquarks, valence quarks, and sea quarks, respectively.

These contributions are determined by the convolution of the diquark and
quark distributions with the fragmentation functions, e.g.,
\begin{equation}
f_{q}^{h}(x_{+},n) = \int_{x_{+}}^{1}
u_{q}(x_{1},n)\cdot G_{q}^{h}(x_{+}/x_{1}) dx_{1}\ \ .
\end{equation}

In the calculation of the inclusive spectra of secondaries produced in
$pA$ collisions we should consider the possibility of one or several Pomeron
cuts in each of the $\nu$ blobs of proton-nucleon inelastic interactions.
For example, in Fig.~1c it is shown one of the diagrams contributing to the
inelastic interaction of a beam proton with two target nucleons. In the
blob of the proton-nucleon1 interaction one Pomeron is cut, while
in the blob of the proton-nucleon2 interaction two Pomerons are cut. The
contribution of the diagram in Fig.~1c to the inclusive spectrum is
\begin{eqnarray}
\frac{x_E}{\sigma_{prod}^{pA}}\cdot\frac{d \sigma}{dx_F} & = & 2\cdot
W_{pA}(2)\cdot w^{pN_1}_1\cdot w^{pN_2}_2\cdot\left\{
f^h_{qq}(x_+,3)\cdot f^h_q(x_-,1)\right. + \nonumber\\ \nonumber & + &
f^h_q(x_+,3)\cdot f^h_{qq}(x_-,1) + f^h_s(x_+,3)\cdot[f^h_{qq}(x_-,2) +
f^h_q(x_-,2) + \\ & + & 2\cdot f^h_s(x_-,2)] \left. \right\} \;,
\end{eqnarray}
where $W_{pA}(2)$ is the probability of interaction with namely two target
nucleons.

It is essential to take into account all digrams with every possible Pomeron
configuration and its corresponding permutations. The diquark and quark distributions and
the fragmentation functions are the same as in the case of $pN$
interaction.

The total number of exchanged Pomerons becomes as large as
\begin{equation}
\langle n \rangle_{pA} \sim
\langle \nu \rangle_{pA} \cdot \langle n \rangle_{pN} \;,
\end{equation}
where $\langle \nu \rangle_{pA}$ is the average number of inelastic
collisions inside the nucleus
(about 4 for heavy nuclei at SpS energies).

The process shown in Fig.~1c satisfies~\cite{Sh3,BT,Weis,Jar} the condition
that the absorptive parts of the hadron-nucleus amplitude are determined by
the combination of the absorptive parts of the hadron-nucleon amplitudes.

In the case of a nucleus-nucleus collision, in the fragmentation region of
projectile we use the approach \cite{Sha,Shab,JDDS}, where the beam of
independent nucleons of the projectile interact with the target nucleus,
what corresponds to the rigid target approximation \cite{Alk} of
Glauber Theory. In the target fragmentation region, on the contrary, the
beam of independent target nucleons interact with the projectile nucleus,
these two results coinciding in the central region. The corrections for energy
conservation play here a very important role if the initial energy is not
very high. This approach was used in \cite{JDDS} for the succsessful
description of $\pi^{\pm}$, $K^{\pm}$, $p$, and $\bar{p}$ produced in Pb-Pb
collisions at 158 GeV per nucleon.

In the present paper we consider the spectra of secondary baryons and
antibaryons, as well as their differences, i.e. the net baryon spectra.
At low energies, the net baryon spectra coincide with the spectra of baryon,
while at asymptotically high energies they are negligible due to the even
signature of the Pomeron trajectory. The energy dependence of the  net baryon
spectra between these two limits strongly depends on the production mechanism.

\subsection{String Junction contribution}

In the string models, baryons are considered as configurations
consisting of three connected strings (related to three valence quarks),
called  String Junction (SJ) \cite{Artru,IOT,RV,Khar}, this picture leading
to some quite general phenomenological predictions.
%Such a baryon
%structure is supported by lattice calculations \cite{latt}.

%This picture leads to some general phenomenological predictions.
The production of a baryon-antibaryon pair in the central region usually
occurs via $SJ$-$\overline{SJ}$ pair production (SJ has upper color indices,
whereas anti-SJ ($\overline{SJ}$) has lower indices), which then combines
with sea quarks and sea antiquarks into a $B\bar{B}$ pair
\cite{RV,VGW}, as it is shown in Fig.~2a.
\begin{figure}[htb]
\centering
%\vskip -2.5cm
%\includegraphics[width=.5\hsize]{fi01.eps}
\includegraphics[width=.5\hsize]{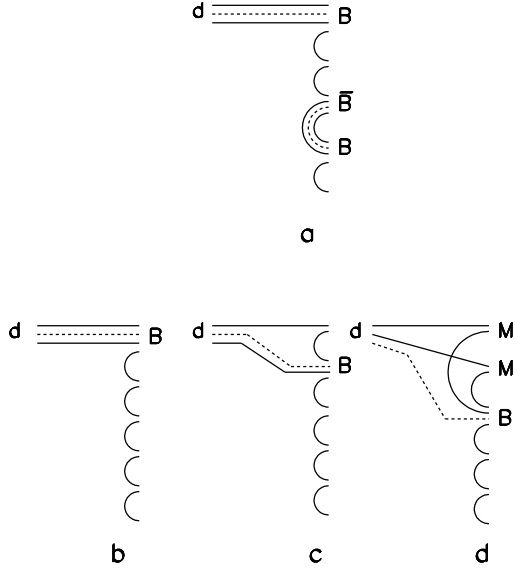}
%\vskip -.3cm
\caption{\footnotesize
QGSM diagrams describing secondary baryon production:
(a) usual $B\bar{B}$ central production with
production of new SJ pair; (b) initial SJ together with two valence
quarks and one sea quark; (c) initial SJ together with one valence
quark and two sea quarks; (d) initial SJ together with three sea quarks.}
\end{figure}

However, in the processes with incident baryons there
exists another possibility to produce a secondary baryon in the central
region, called SJ diffusion. The quantitative description of the baryon number
transfer due to SJ diffusion in rapidity space was obtained in \cite{ACKS}
and following papers \cite{AMPS,MPS,BS,AMS,Olga,MRS}.

In the QGSM the differences in the spectra of secondary baryons and
antibaryons appear for processes which
present SJ diffusion in rapidity space. These differences only vanish rather
slowly when the energy increases.

To obtain the net baryon charge, and according to ref.~\cite{ACKS}, we consider
three different possibilities. The first one is the fragmentation of the
diquark giving rise to a leading baryon (Fig.~2b). A second possibility is
to produce a leading meson in the first break-up of the string and a baryon
in a subsequent break-up~\cite{Kai,22r} (Fig.~2c). In these two first
cases the baryon number transfer is possible only for short distances
in rapidity. In the third case, shown in Fig.~2d, both initial valence
quarks recombine with sea antiquarks into mesons, $M$, while a secondary
baryon is formed by the SJ together with three sea quarks~\cite{MPS,ACKS,KP1}.

The fragmentation functions for the secondary baryon $B$ production
corresponding to the three processes shown in Fig.~2b, 2c, and 2d  can be
written as follows (see~\cite{ACKS} for more details):
\begin{eqnarray}
G^B_{qq}(z) &=& a_N\cdot v^B_{qq} \cdot z^{2.5} \;, \\
G^B_{qs}(z) &=& a_N\cdot v^B_{qs} \cdot z^2\cdot (1-z) \;, \\
G^B_{ss}(z) &=& a_N\cdot\varepsilon\cdot v^B_{ss} \cdot z^{1 - \alpha_{SJ}}
\cdot (1-z)^2  \;,
\end{eqnarray}
where $a_N$ is the normalization
parameter, and $v^B_{qq}$, $v^B_{qs}$, $v^B_{ss}$ are the relative
probabilities for different baryons production that can be found by simple
quark combinatorics \cite{AnSh,CS}. These probabilities depend on the
strangeness suppression factor $S/L$, and we use $S/L = 0.32$ following 
\cite{AKMS}.
 
The contribution of the graph in Fig.~2d has in QGSM a coefficient
$\varepsilon$ which determines the small probability for such a baryon
number transfer.

The fraction $z$ of the incident baryon energy carried by the secondary
baryon decreases from Fig.~2b to Fig.~2d.
Only the processes in Fig.~2d can contribute to the inclusive spectra in the
central region at high energies if the value of the intercept of the SJ
exchange Regge-trajectory, $\alpha_{SJ}$, is large enough. The analysis in
\cite{MRS} gives a value of $\alpha_{SJ} = 0.5 \pm 0.1$, that is in agreement
with the ALICE Collaboration result, $\alpha_{SJ} \sim 0.5$ \cite{ALICE}, obtained at
LHC. In the calculations of these effects we use the following values of the
parameters \cite{MRS}:
\begin{equation}
\alpha_{SJ}\, =\, 0.5\;\; {\rm and} ~\varepsilon\, =\, 0.0757\,.
\end{equation}

\subsection{Contribution from interaction with clusters}

In the case of interaction with a nuclear target some secondaries can be
produced in the kinematical region forbidden for the interaction with a free
nucleon. Such processes are called the cumulative ones, the simplest example
being the production of secondary nucleons in the backward
hemisphere in the laboratory frame.

Usually, the cumulative processes are considered as a result of the coherent
interaction of a projectile with a multiquark cluster, i.e. with a group of
several nucleons which are at short distances from each other that appears as
as a fluctuation of the nuclear matter \cite{SF,Blokh}.

The inclusive spectra of the secondary hadron $h$ in the central region is
determined at high energies by double-Pomeron diagrams \cite{AKM}.
The case of $pp$ collision is shown in Fig.~3a. In the case of proton-nucleus
collisions two different possibilities exist, the interactions with individual target
nucleons (Fig.~3b) and the secondary production on cluster (Fig.~3c). For
nucleus-nucleus collisions there are four possibilities (Figs.~3d, 3e,
3f, and 3g). The one in Fig.~3g corresponds to a new process where a secondary
hadron is produced by the interaction of two clusters.
\begin{figure}[htb]
\centering
%\vskip -2.5cm
\includegraphics[width=.5\hsize]{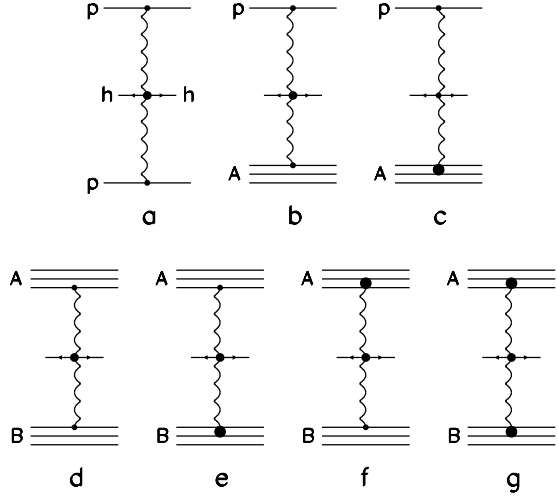}
%\vskip -.3cm
\caption{\footnotesize
Reggeon diagrams for the different possibilities corresponding to
the inclusive spectra of a secondary hadron $h$ produced in the
central region in (a) $pp$, (b,c) $p$A, and (d$-$g) AB collisions. Pomerons are
shown by wavy lines.}
\end{figure}

It was shown in refs. \cite{Efr,Efre} that in the case of secondary production from
the cluster fragmentation, the inclusive spectra can be calculated in the
framework of the QGSM with the same quark and diquark fragmentation
functions. The only difference comes from the quark and diquark distributions,
$u_q^{cl}(x,n,k)$ and $u_{qq}^{cl}(x,n,k)$, where $k$ is the number of nucleons
in the cluster. The  distributions $u_i^{cl}(x,n,k)$ can also be calculated
by using the Reggeon counting rules, and they take the form:
\begin{eqnarray}
u^{cl}_{uu}(x,n) & = & C_{uu}\cdot x^{\alpha_R-2\alpha_B+1}
\cdot(k-x)^{-\alpha_R N +m_1}\nonumber \\ \nonumber
u^{cl}_{ud}(x,n) & = & C_{ud}\cdot x^{\alpha_R-2\alpha_B}\cdot(k-x)^{-\alpha_R +N+m_2}
\\ \nonumber
u^{cl}_u(x,n,k) & = & C_u\cdot x^{-\alpha_R}\cdot(k-x)^{\alpha_R-2\alpha_B + N +m_2}
\\ \nonumber
u^{cl}_d(x,n) & = & C_d\cdot x^{-\alpha_R}\cdot(k-x)^{\alpha_R-2\alpha_B+1 + N+m_1}
\\
u^{cl}_s(x,n) & = & C_d\cdot x^{-\alpha_R}\cdot(1-x)^{\alpha_R-2\alpha_B+(n-1)+d} \;,
\end{eqnarray}
where
\begin{equation}
N  =  2\cdot(1-\alpha_B)\cdot(k - 1) \;.
\end{equation}
The function $u^{cl}_s(x,n)$ does not depend on $k$. All these functions
$u^{cl}_i(x,n,k)$ are normalized to unity and the values $m_1$ and $m_2$ can
be found from conditions in Eq.~(2).

This approach was succesfully used in \cite{Efr,Efre} for the description of
cumulative particles produced in $hA$ collisions. In the present paper we use
it for describing the enhancement of strangeness production on nuclear
targets in the central region.

The probability to find a proton in the backward hemisphere in high energy $pA$
collisions reach values up to 10\%. Keeping in mind that it is only a part
of cumulative processes, in the numerical calculations and for every
hadron-nucleon interaction we take the probability to interact with a cluster
$V^{cl}$, and the probability to actually interact with a nucleon $1 - V^{cl}$.
We use the numerical values:
\begin{equation}
V^{cl} = 0.2 \; {\rm (for~medium A)}\;, V^{cl} = 0.3 \; {\rm (for ~heavy A)}\;,
\end{equation}
as the maximal reasonable values of $V^{cl}$.

\subsection{\bf Inelastic screening (percolation) effects}

The QGSM gives a reasonable description \cite{MPS,KTMS,JDDS,Sh4} of the 
inclusive spectra of different secondaries produced both in hadron-nucleus and 
in nucleus-nucleus collisions at energies $\sqrt{s_{NN}}$ = 14$-$30 GeV.  

At RHIC energies the situation drastically changes. The spectra of
secondaries produced in $pp$ collisions are described rather well 
\cite{MPS}, but the RHIC experimental data for Au+Au collisions 
\cite{Phob,Phen} give clear evidence of the inclusive density suppression 
effects which reduce by a factor $\sim$0.5 the midrapidity inclusive density,
when compared to the predictions based on the superposition picture 
\cite{CMT,Sh6,AP}. This reduction can be explained by the inelastic screening
corrections connected to multipomeron interactions~\cite{CKTr} (see Fig.~4). 

%The effect is
%very small for integrated cross sections (many of them are determined only by
%geometry), but it is very important~\cite{CKTr} for the calculations of 
%secondary multiplicities and inclusive densities at high energies.
\begin{figure}[htb]
\centering
\vskip -1.5cm
\includegraphics[width=.42\hsize]{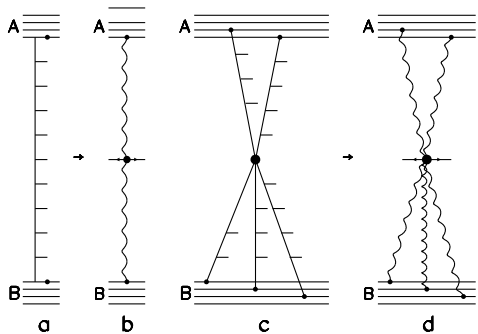}
%\vskip -1.7cm
%\includegraphics[width=.5\hsize]{fi00a.eps}
%\vskip -.8cm
\caption{\footnotesize
(a) Multiperipheral ladder corresponding to the inclusive cross section of
diagram (b),  and (c) fusion of several ladders corresponding to the inclusive
cross section of diagram (d).}
\end{figure}

At energies $\sqrt{s_{NN}} \leq $ 30$-$40 GeV, the inelastic processes are 
determined by the production of one (Fig.~4a) or several (Fig.~4c) multiperipheral
ladders, and the corresponding inclusive cross sections are described by the diagrams
of Fig.~4b and Fig.~4d. 

In accordance with the Parton Model~\cite{Kan,NNN}, the fusion of multiperipheral 
ladders shown in Fig.~4c becomes more and more important with the increase of the
energy, resulting in the reduction of the inclusive density of secondaries. Such
processes correspond to the enhancement Reggeon diagrams of the type of Fig.~4d,
and to even more complicate ones. All these diagrams are proportional to the squared
longitudinal form factors of both colliding nuclei \cite{CKTr}, so their contribution
becomes negligible when the energy decreases. Following the estimations presented
in reference~\cite{CKTr}, the RHIC energies are just of the order of magnitude
needed to observe this effect. 

%The 
%inelastic screening can make \cite{CKTr} the inclusive density to decrease in 
%the midrapidity region about two times at RHIC energies and about three times 
%at LHC energies, with respect to the calculations without inelastic screening.
However, all quantitative estimations are model dependent. The numerical weight of the 
contribution of the multipomeron diagrams is rather unclear due to the many 
unknown vertices in these diagrams. The number of unknown parameters can be 
reduced in some models, and, as an example, in reference~\cite{CKTr} the 
Schwimmer model~\cite{Schw} was used for the numerical estimations.
Also, in~\cite{Ost} the phenomenological multipomeron vertices of eikonal 
type were introduced for the summation of the enhancement diagram. 

%Another approache was used in~\cite{Ost}, where the phenomenological 
%multipomeron vertices of eikonal type were introduced for enhancement 
%diagram summation. 
Another possibility to estimate the contribution of the
diagrams with Pomeron interaction comes~\cite{JUR,JUR1,BP,JDDSh,BJP}
from Percolation Theory. The percolation approach and its previous version, 
the String Fusion Model~\cite{SFM,SFM1,SFM2}, predicted the multiplicity
suppression seen at RHIC energies, long before any RHIC data were measured.

New calculations of inclusive densities and multiplicities in percolation 
theory both in $pp$~\cite{CP1,CP2}, and in heavy ion collisions~\cite{CP2,CP3},
are in good agreement with the experimental data in a wide energy region.

%The percolation model also provides a reasonable description of the 
%transverse momentum distribution (at low and intermediate $p_T$) including 
%the Cronin effect and the behaviour of the baryon/meson ratio
%\cite{Dias,Paj,Paj1}. Most of the effects predicted by percolation can be 
%seen as a direct consequence of the strong colour fields produced in the 
%collision. This feature is common to other approaches as the Colour Glass 
%Condensate~\cite{McL,Ele}.

%, where a $p_T$ scaling is also obtained \cite{Schaff}. 
In order to account for the percolation (inelastic screening) effects in the 
QGSM, it is technically more simple~\cite{MPS1} to consider the maximal number
of Pomerons $n_{max}$ emitted by one nucleon in the central region that 
can be cut. These cut Pomerons lead to the different final states. Then 
the contributions of all diagrams with $n \leq n_{max}$ are accounted for as 
at lower energies. The larger number of Pomerons $n > n_{max}$ can also be 
emitted obeying the unitarity constraint, but due to the fusion in the final state 
(at the quark-gluon string stage), the cut of $n > n_{max}$ Pomerons results 
in the same final state as the cut of $n_{max}$ Pomerons.

By doing this, all model calculations become very similar to the percolation
approach. The QGSM fragmentation formalism allows one to 
calculate the integrated over $p_T$ spectra of different secondaries as the
functions of rapidity and $x_F$.

In this frame, we obtain a reasonable agreement with the experimental data 
on the inclusive spectra of secondaries produced in d+Au collisions 
at RHIC energy \cite{MPS1} with a value $n_{max} = 13$, and in p+Pb collisions
at LHC energy \cite{pPbold} with the value $n_{max} = 23$. It has been 
shown in~\cite{JDDCP} that the number of strings that can be used for the secondary
production should increase with the initial energy. 

%Thus, in the following calculations we use the value $n_{max} = 13$
%at the LHC energy $\sqrt{s}$ = 5 TeV, that can be considered as the 
%normalization to ALICE data for all charged secondaries. The predictive 
%power of our calculation applies for different sorts of secondaries in 
%midrapidity region, as well as the calculations in the fragmentation region.

%In the following calculations we use the simplest version 
%of this approach, in which we neglect the energy dependence of $n_{max}$,
%i.e. we use a fixed $n_{max} = 13$ at LHC energy (see discussion in the
%section Conclusions).

\section{Numerical results}

\subsection{Net baryon spectra}
One example of the QGSM description of the $x_F$-spectra of secondary protons
and antiprotons measured in $pp$ collisions at 158 GeV/c by NA49 Collaboration
\cite{NA49pp} is presented in Fig.~5. 
\begin{figure}[htb]
\centering
%\vskip -2.5cm
\includegraphics[width=.6\hsize]{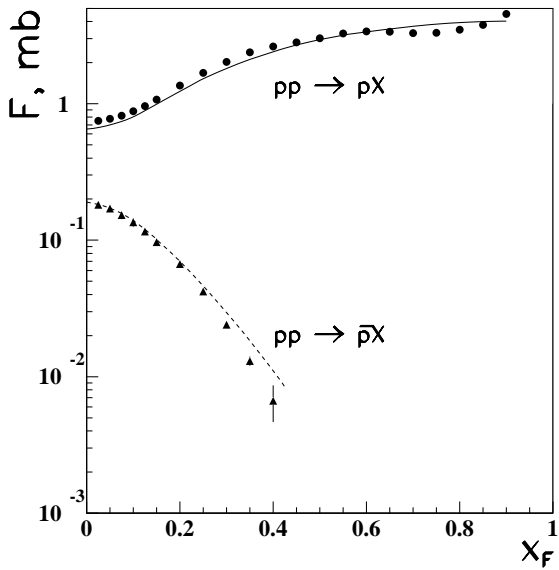}
\vskip -.6cm
\caption{\footnotesize
QGSM $x_F$-spectra of secondary protons and antiprotons produced in $pp$ collisions
at 158 GeV/c compared to the experimental data by the NA49 Collaboration~\cite{NA49pp}.}
\end{figure}

The QGSM results for net proton ($p - \bar{p}$) and net $\Lambda$-hyperon
($\Lambda - \bar{\Lambda}$) productions at 200 GeV per nucleon
are compared to the experimental data by the NA35 Collaboration~\cite{NA35} 
in Figs.~6 and 7.
\begin{figure}[htb]
\centering
\vskip -1.cm
\includegraphics[width=.4\hsize]{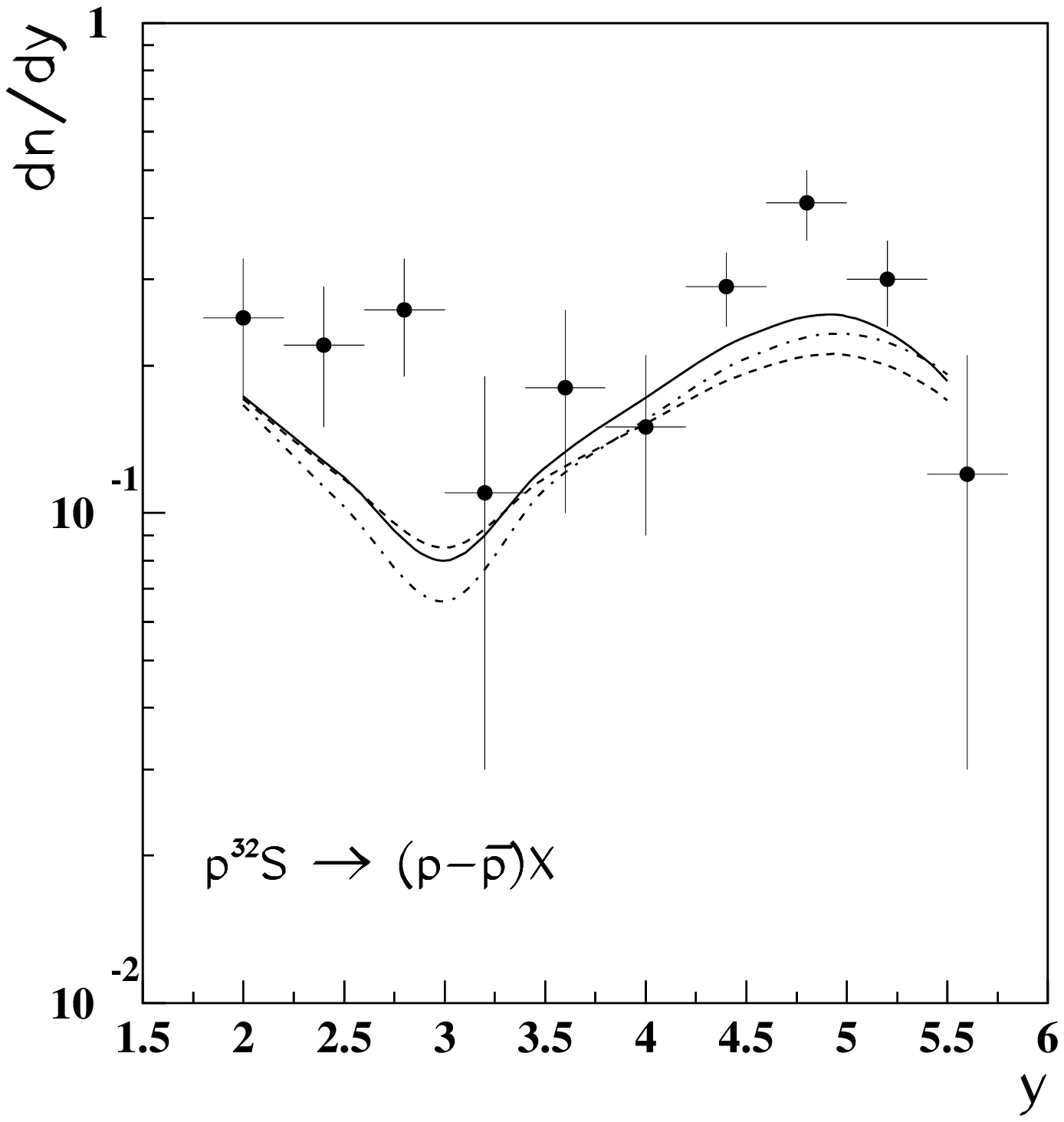}
\includegraphics[width=.4\hsize]{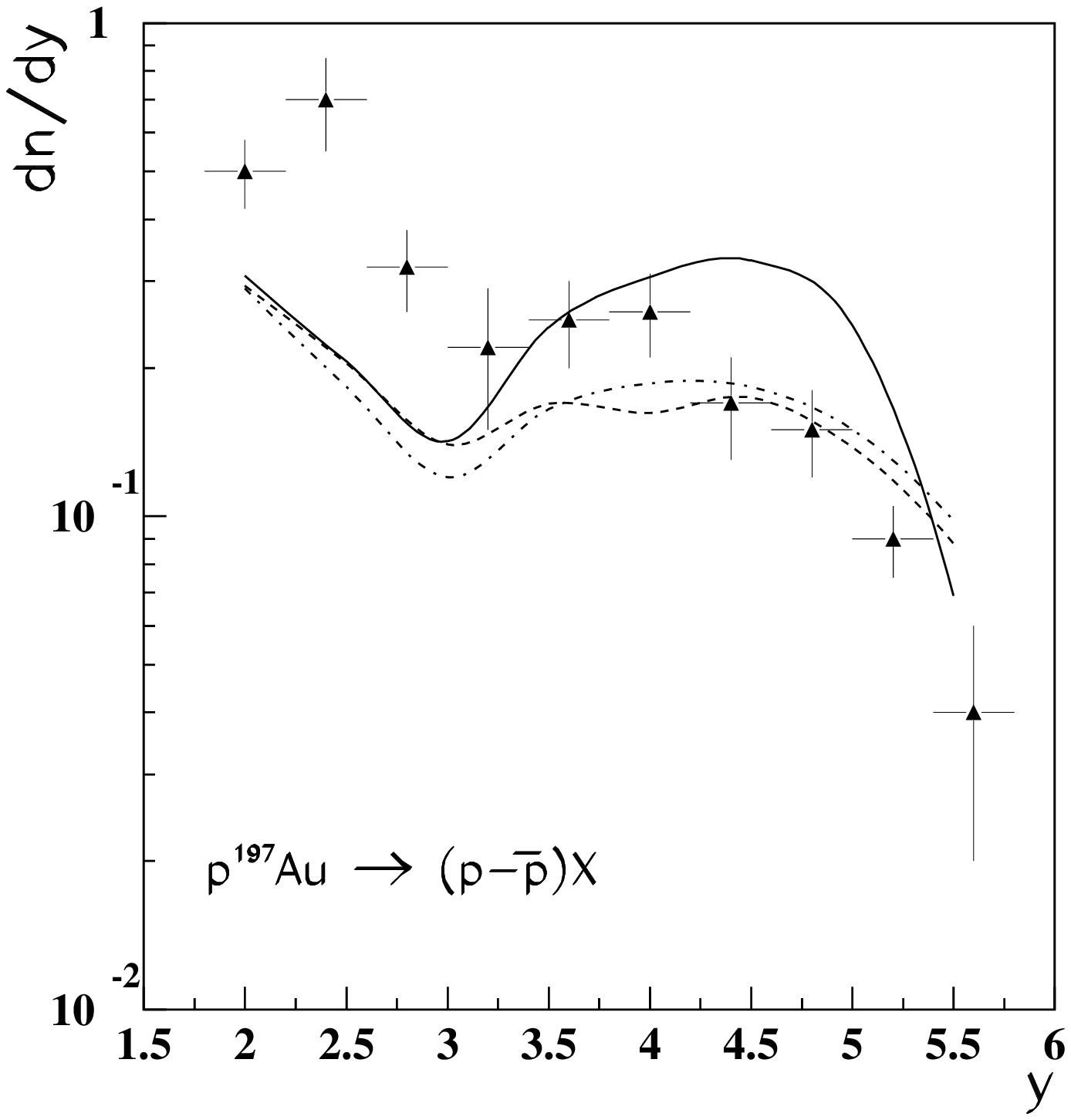}
\includegraphics[width=.4\hsize]{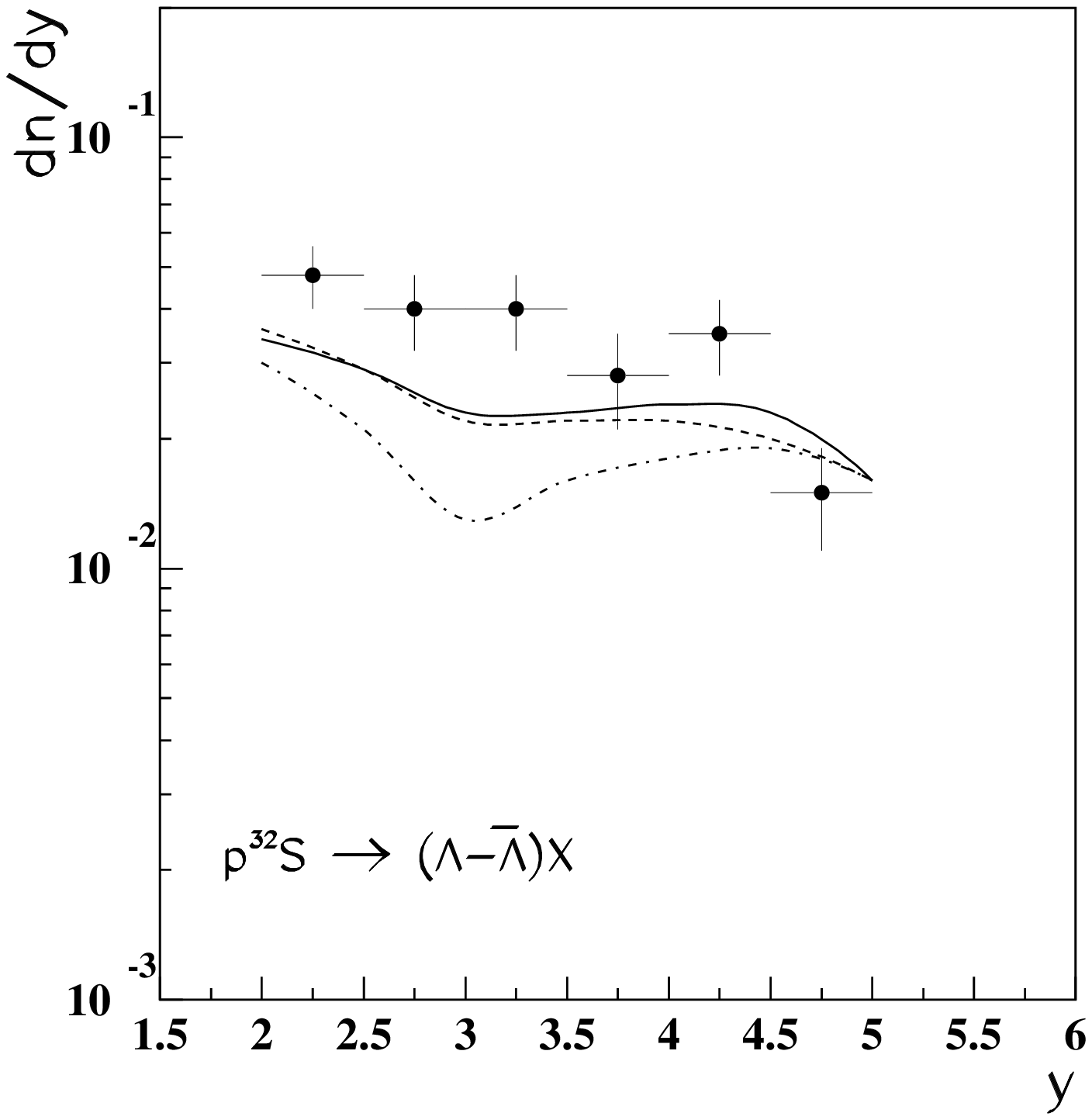}
\includegraphics[width=.4\hsize]{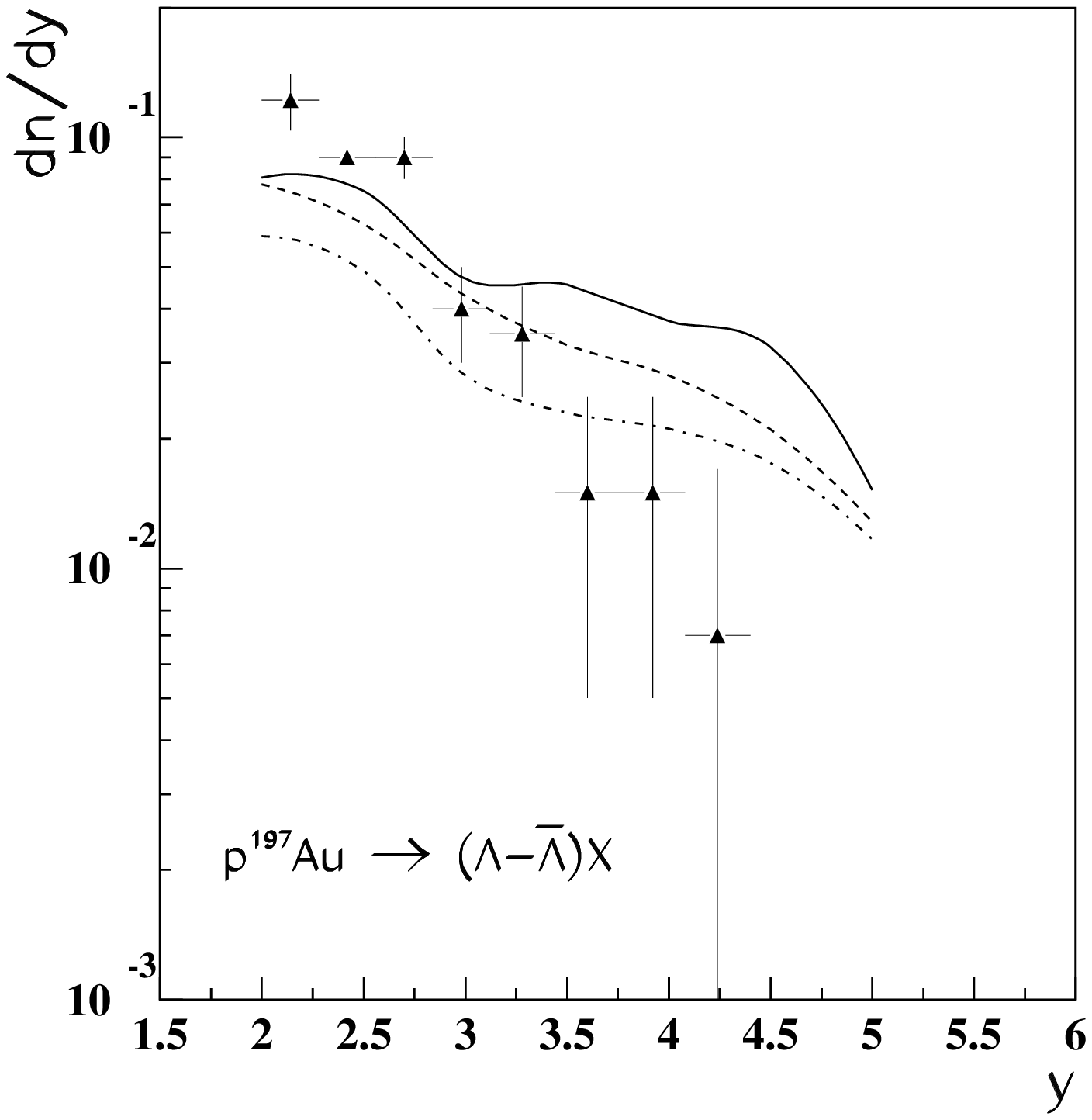}
\vskip -.3cm
\caption{\footnotesize
Net proton $p - \bar{p}$ (upper panels) and net $\Lambda$-hyperon
$\Lambda - \bar{\Lambda}$ (lower panels) production in $p-^{32}$S (left
panels) and $p-^{197}$Au (right panels) collisions at 200 GeV per nucleon.
Solid curves show the QGSM calculations with both SJ and cluster
contributions, dashed curves with SJ contributions but without the cluster ones,
and dotted curves without both SJ and cluster contributions.}
\end{figure}

The data for net baryon production by proton beam interaction with 
 $^{32}$S  and $^{197}$Au nuclear targets are presented in Fig.~6 in the 
central and beam fragmentation regions as function of rapidity in the 
laboratory system. The absolute normalization of  $dn/dy$ in all cases is 
determined by the data of proton and antiproton production in $pp$ collisions 
at similar energies.

The results of the QGSM calculations without SJ and cluster contributions are
shown in Fig.~6 by dotted lines. Dashed lines show the same calculations 
with SJ contributions but without the cluster ones, and solid lines show the
results with both SJ and cluster contributions. The corrections for very high
energy interactions described in Subsection 2.4 are negligible at this energy.

In the case of net proton production in $p-^{32}$S collisions, all three
curves are close to each other and they are in reasonable agreement with the
experimental data. In the case of $p-^{197}$Au collisions the number of net protons
is too small at small rapidities, what can be explained by the influence of the target  
fragmentation region. The nuclear cluster contribution, which is important
mainly in the beam fragmentation region, seems to be too large.
\begin{figure}[htb]
\centering
\vskip -1.cm
\includegraphics[width=.4\hsize]{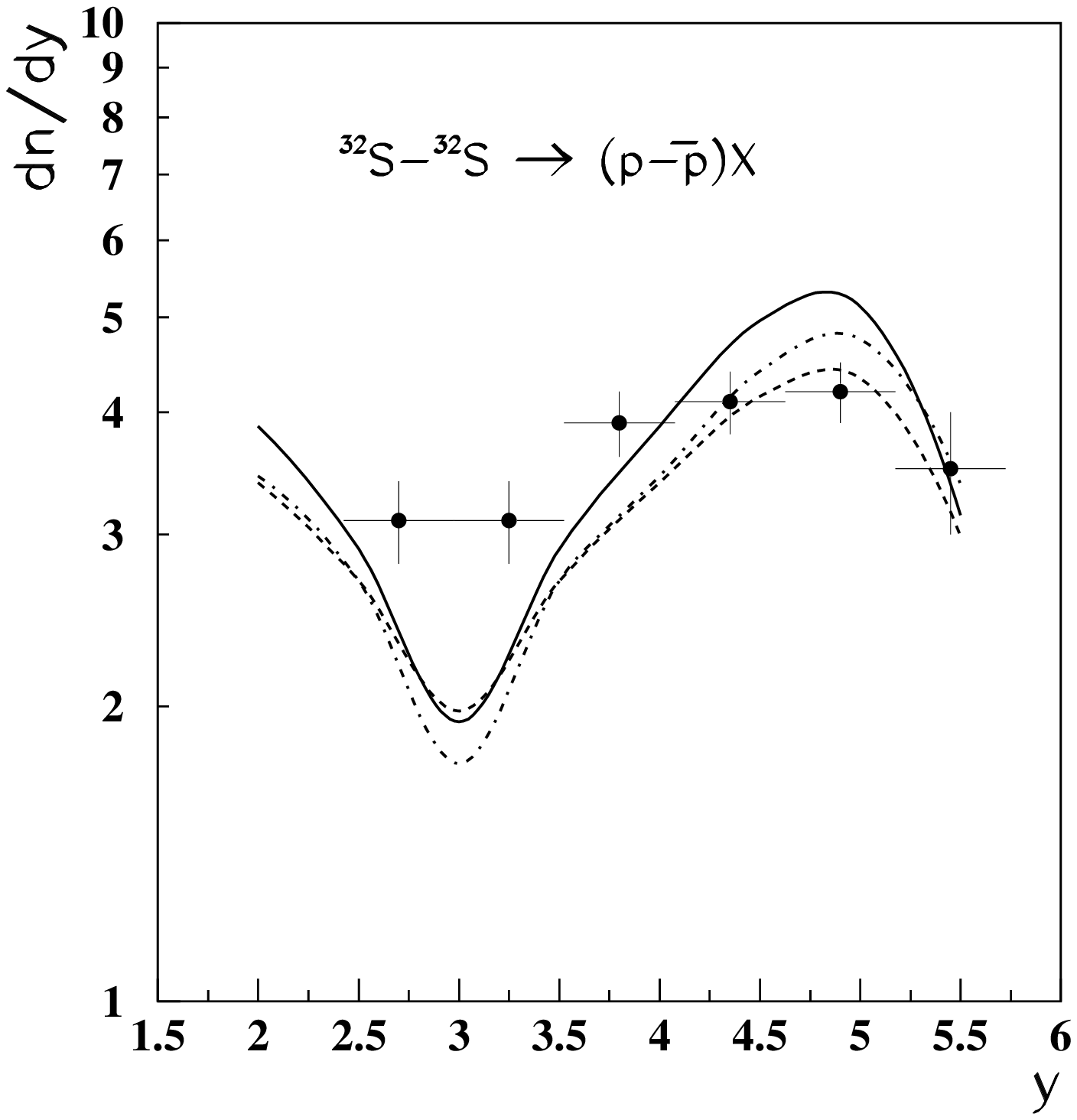}
\includegraphics[width=.4\hsize]{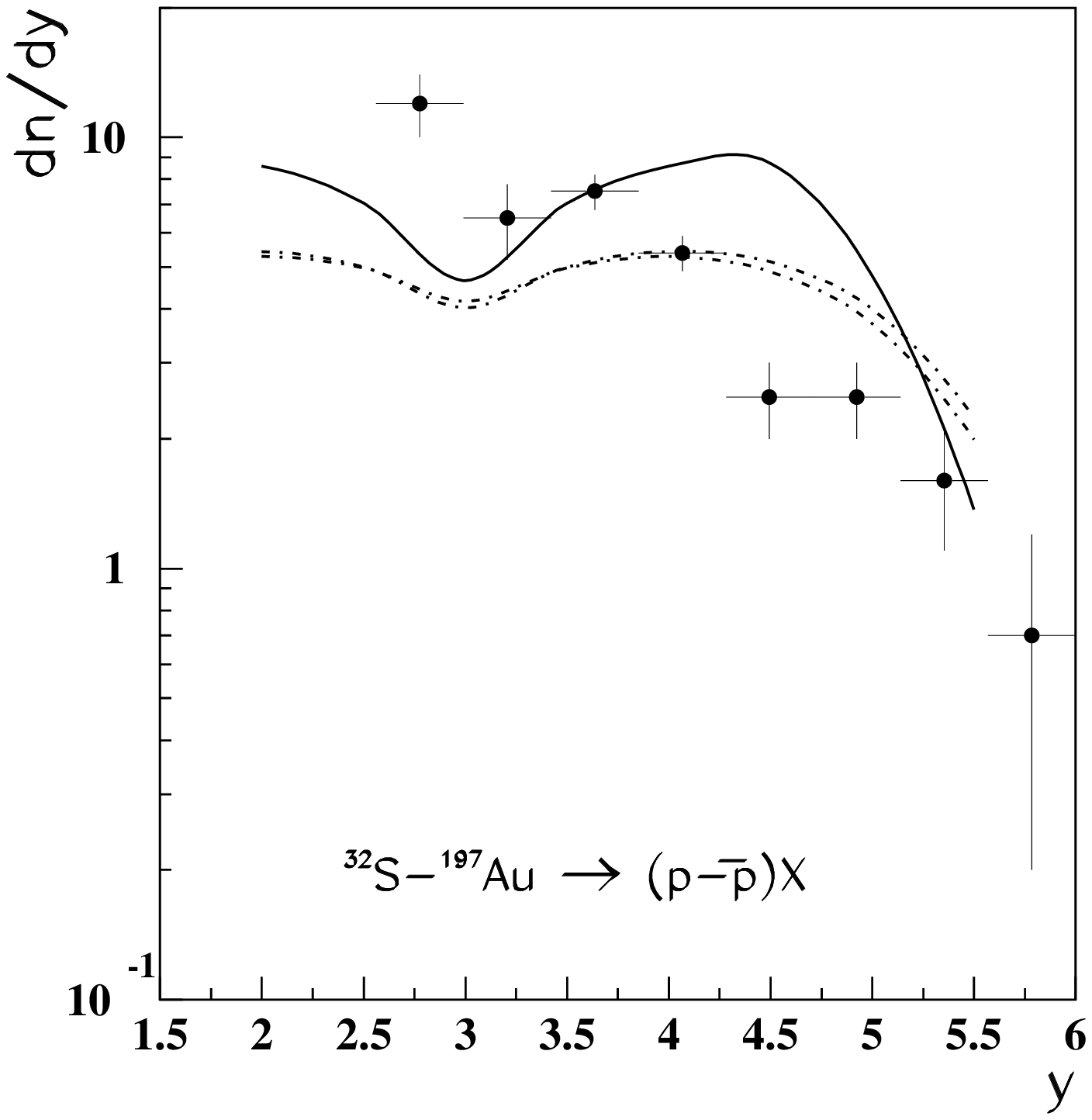}
\includegraphics[width=.4\hsize]{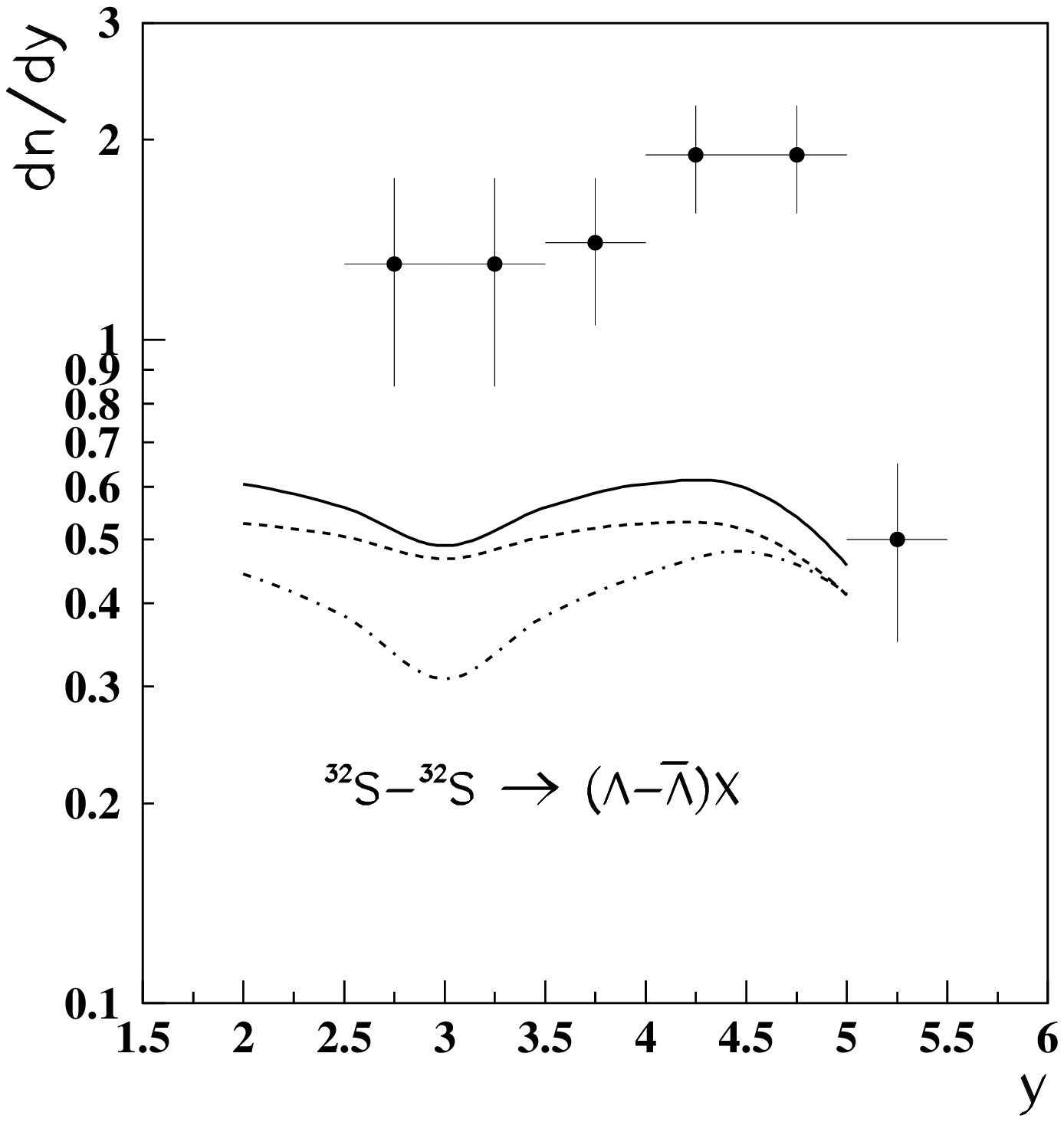}
\includegraphics[width=.4\hsize]{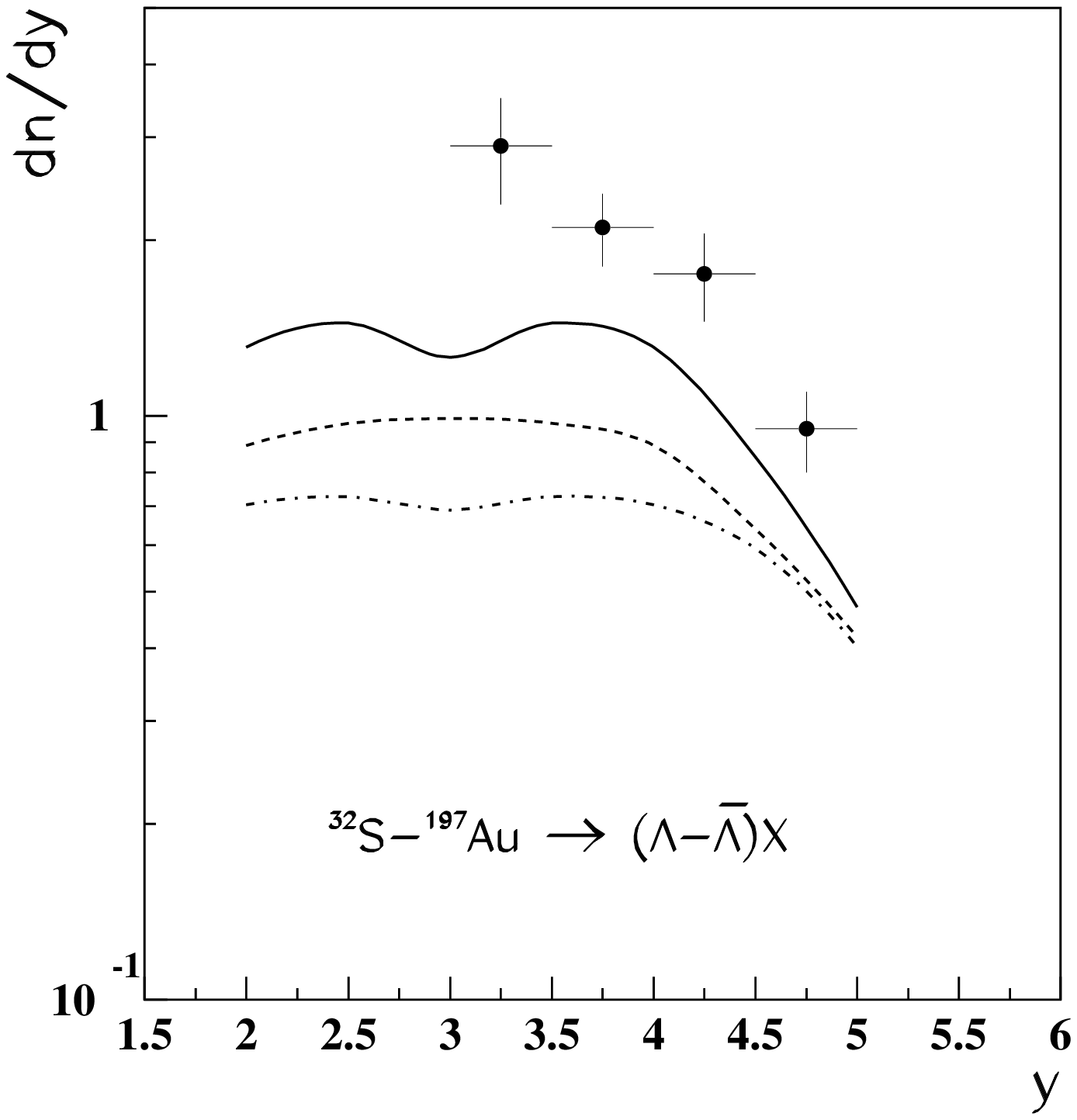}
\vskip -.3cm
\caption{\footnotesize
Net proton $p - \bar{p}$ (upper panels) and net $\Lambda$-hyperon
$\Lambda - \bar{\Lambda}$ (lower panels) production in $^{32}$S - $^{32}$S
(left panels) and in $^{32}$S - $^{197}$Au (right panels) collisions at
200 GeV per nucleon. Solid curves show the QGSM calculations with both SJ
and cluster contributions, dashed curves with SJ contributions but without
the cluster ones, and dotted curves without both SJ and cluster contributions.}
\end{figure}

In  the case of net $\Lambda$-hyperon production the experimental error bars
are rather large, and one can talk of general semiquantitative agreement of the
QGSM calculations with the data in the lower panels of Fig.~6.

For the sulphur $^{32}$S beam shown in Fig.~7, the theoretical calculations
are in reasonable agreement with the data for net proton
production, whereas the data on net $\Lambda$-hyperon production are systematicaly
higher than all calculated curves. In any case, though, all disagreements are not
large than $\sim$30\%.

The data of the NA35 Collaboration \cite{NA35} are compatible with the results
of the NA44 Collaboration \cite{NA44}.
%\end{document}

\subsection{$\Lambda$ and $\bar{\Lambda}$ production in midrapidity region}

The NA49 Collaboration obtained experimental data \cite{NA49a} for
yields of $\Lambda$ and $\bar{\Lambda}$ hyperons in midrapidity region
($\vert y \vert < 0.4$) in the central C+C, Si+Si, and Pb+Pb collisions 
(5\% centrality), at 158 GeV per nucleon. These results 
are presented in Table~1, together with the QGSM results obtained
for the same rapidities and centralities.
%but the calculated yields of $\Xi^-$, 
%and $\overline{\Xi^+}$ are significantly smaller than the NA49 Collaboration 
%data. This disagreement can be very important and we will discuss it in more
%details in next Subsection.
%\newpage

\begin{center}
\vskip 5pt
\begin{tabular}{|c||c|c|c|c|c|c|} \hline

$\sqrt{s}$ (GeV) & Reaction   & QGSM & Experiment  dn/dy  \\
%\hline
  \hline

17.2 & C +C $\to \Lambda$ & 0.237 & $0.24 \pm 0.01 \pm 0.04$, 
\cite{NA49a}  \\

& C + C $\to \bar{\Lambda}$ & 0.064 & $0.064 \pm 0.003 \pm 0.010$ 
\cite{NA49a} \\ \hline
 \hline

17.2 & Si + Si $\to \Lambda$ & 0.69 & $0.68 \pm 0.04 \pm 0.13$, 
\cite{NA49a} \\

& Si + Si $\to \bar{\Lambda}$ & 0.17 & $0.16 \pm 0.007 \pm 0.038$, 
\cite{NA49a} \\  \hline

17.2 & Pb + Pb $\to \Lambda$ & 9.4 & $12.9 \pm 0.7 \pm 1.5$ 
\cite{NA49a} \\

& Pb + Pb $\to \bar{\Lambda}$ & 2.05 &
$1.4 \pm 0.3 \pm 0.2$ 
\cite{NA49a} \\ \hline

17.2 & p + Be $\to \Lambda$ & 0.034  & $0.034 \pm 0.0005 \pm 0.003$ 
\cite{NA57} \\

(m.b.) & p + Be $\to \bar{\Lambda}$ & 0.010 &
$0.011 \pm 0.0002 \pm 0.001$ 
\cite{NA57} \\ \hline

17.2 & p + Pb $\to \Lambda$ & 0.074& $0.060 \pm 0.002 \pm 0.006$ 
\cite{NA57} \\

(m.b.) & p + Pb $\to \bar{\Lambda}$ & 0.0019 &
$0.015 \pm 0.001 \pm 0.002$ 
\cite{NA57} \\ \hline

17.2 & Pb + Pb $\to \Lambda$ & 9.4 & $18.5 \pm 1.1 \pm 1.8$ 
\cite{NA57} \\

& Pb + Pb $\to \bar{\Lambda}$ & 2.05 &
$2.44 \pm 0.14 \pm 0.24$ 
\cite{NA57} \\ \hline

62.4 & Au + Au $\to \Lambda$ & 11.1 & $15.7 \pm 0.3\pm2.3 $ 
\cite{STAR} \\

& Au + Au $\to \bar{\Lambda}$ & 8.2 &
$8.3 \pm 0.2\pm1.1 $ 
\cite{STAR} \\ \hline

200 & Cu + Cu $\to \Lambda$ & 3.82 & $4.68 \pm 0.45 $ 
\cite{STAR1} \\

& Cu + Cu $\to \bar{\Lambda}$ & 3.34 &
$3.79 \pm 0.37 $ 
\cite{STAR1} \\ \hline

200 & Au + Au $\to \Lambda$ & 14.2 & $14.8 \pm 2.4 $ 
\cite{STAR1} \\

& Au + Au $\to \bar{\Lambda}$ & 12.1 &
$11.7 \pm 0.9 $ 
\cite{STAR1} \\ \hline

3000 & Pb + Pb $\to \Lambda$ & 36.2 & - \\

& Pb+Pb $\to \bar{\Lambda}$ & 35.6 & - \\ \hline

\hline\end{tabular}
\end{center}
Table 1. Experimental NA49 \cite{NA49a}, NA57 \cite{NA57}, and STAR \cite{STAR,STAR1}
data for $\Lambda$ and $\bar{\Lambda}$ production at 158GeV per nucleon, and at STAR energies,
and the corresponding description by the QGSM.

%\newpage

%{\Large \bf

%\subsection{NA57 data}
On the other hand, the NA57 Collaboration obtained the experimental data \cite{NA57}
for $\Lambda$ and $\bar{\Lambda}$ yields in midrapidity region $\vert y \vert < 0.5$
in the minimum bias $p$+Be and $p$+Pb interactions, and in central (5\% centrality)
Pb+Pb collisions at 158 GeV per nucleon.

Unfortunately, the data by the NA49 and NA57 Collaborations are not compatible,
as one can see from Table 1, where the values of $dn/dy$ for different hyperons measured by one
collaboration are far outside the error bars of the corresponding values published by the other
collaboration for the same centrality. This is probably due to different experimental event selection.
.
%{\Large \bf

Here again one can see that the calculated yields of $\Lambda$ and
$\bar{\Lambda}$ are in agreement with experimental data on the level
of 20$-$30\% accuracy. 
%Inclusive densities of $\overline{\Xi^+}$
%hyperons are reproduced reasonably for the cases of p+Be and p+Pb collisions
%but are underestimated several times in the case of central Pb+Pb
%interactions. In the case of $\overline{\Omega^+}$ production in   central 
%Pb+Pb collisions the disagreement is more than an order of magnitude.

Hyperon production at higher energies in midrapidity region was also measured at 
RHIC. The data by the STAR Collaboration \cite{STAR,STAR1} for Au + Au and Cu+Cu
collisions at $\sqrt{s_{NN}}$ = 62.4 GeV and 200 Gev are presented in Table 1.
We also give in Table 1 the QGSM predictions for central Pb+Pb collisions at the 
LHC energy $\sqrt{s_{NN}}$ = 3 TeV.

\section{Conclusion}

The QGSM provides a reasonable description of nucleon and $\Lambda$, as well as
their antiparticles, production in 
nucleon-nucleus and nucleus-nucleus collisions at high energies. The level of
numerical accuracy is of about 20$-$30\%. Part of the uncertainty is connected to
discrepancies among the different experimental data.

Inclusive densities of $\overline{\Xi^+}$
hyperons are reasonably reproduced for the cases of p+Be and p+Pb collisions
\cite{NA57}, but they are several times underestimated in the case of central Pb+Pb
interactions \cite{NA57}. For $\overline{\Omega^+}$ production in central Pb+Pb
collisions the disagreement is larger than one order of magnitude.
The physical reasons for these observed disagreements will be discussed in a separate 
paper.

{\bf Acknowledgements}

We are grateful to \frame{A.B. Kaidalov} for useful discussionsi and
comments. This paper was supported by Ministerio de Educaci\'on y Ciencia of
Spain under the Spanish Consolider-Ingenio 2010 Programme CPAN (CSD2007-00042)
and project FPA 2005--01963, by Xunta de Galicia and, in part,
by grant RSGSS-3628.2008.2, and by the State Commitee of Science 
of the Republic of Armenia, through Grant-11-1C015.

\end{document}